\documentclass{easychair}

\pdfoutput=1

\usepackage[T1]{fontenc}
\usepackage{graphicx}
\usepackage{amsmath}
\usepackage{amssymb}
\usepackage{hyperref}
\usepackage{marvosym}
\usepackage{booktabs}
\usepackage{comment}
\usepackage{multicol}
\usepackage{todonotes}
\usepackage{amsthm}
\usepackage{xcolor}
\usepackage{paralist}

\newtheorem*{thm}{Theorem}

\renewcommand{\geq}{\geqslant}
\renewcommand{\leq}{\leqslant}
\newcommand{\limp}{\Rightarrow}

\newcommand{\dpi}{\mathrm{\Pi}}
\newcommand{\tp}{\mathsf{tp}}

\title{Experiments with Choice\\ in Dependently-Typed Higher-Order Logic}
\titlerunning{Choice in Dependently-Typed Higher-Order Logic}

\author{Daniel Ranalter\inst{1} \and Chad E. Brown\inst{2} \and Cezary Kaliszyk\inst{1}}
\authorrunning{Ranalter, Brown, \and Kaliszyk}
\institute{
  University of Innsbruck, Innsbruck, Austria\\
  \email{\{daniel.ranalter,cezary.kaliszyk\}@uibk.ac.at}
  \and
  Czech Technical University in Prague, Prague, Czech Republic 
}

\def\vdashs{\vdash^{\mathsf{s}}}
\def\vdashd{\vdash^{\mathsf{d}}}
\def\vdashschoice{\vdash^{\mathsf{s}\varepsilon}}
\def\vdashdchoiceone{\vdash^{\mathsf{d}\varepsilon_1}}
\def\vdashdchoicetwo{\vdash^{\mathsf{d}\varepsilon_2}}

\usepackage{pifont}
\newcommand{\cmark}{\textcolor[rgb]{0,.5,0}{\ding{51}}}%
\newcommand{\xmark}{\tiny\raisebox{.5mm}{\ding{55}}}%

\begin{document}

\maketitle

\begin{abstract}
Recently an extension to higher-order logic ---
called DHOL --- was introduced, enriching the language with dependent
types, and creating a powerful extensional type theory.
In this paper we propose two ways how choice can be added to DHOL.
We extend the DHOL term structure by 
Hilbert's indefinite choice operator \(\epsilon\), define a translation
of the choice terms to HOL choice that extends the existing translation
from DHOL to HOL
and show that the extension of the translation is complete and give an
argument for soundness.
We finally evaluate the extended translation on a set of dependent HOL problems
that require choice.
\end{abstract}

\section{Introduction}
Dependently-typed higher-order logic (DHOL), introduced by Rothgang et al.~\cite{RRB23ext}, is an extension of classical 
higher-order logic as originally presented by Church~\cite{C40}, albeit with a few modifications. DHOL turns the 
simply-typed lambda calculus with a base type for Booleans and an equality predicate for every type into a 
dependently-typed and extensional type theory. This makes it possible to define, for example, fixed-length lists or the 
type of fixed-size finite sets, by replacing the simple function type \(A\rightarrow B\) with its dependent version \(\Pi x:A.B\). 
While the extensionality comes at the cost of undecidable type-checking, initial experiments by Niederhauser et 
al.~\cite{NBK24} suggest that on the other side of this trade-off lies the ability to find proofs of otherwise 
unobtainable problems. To make DHOL usable by a wider array of theorem provers,
Rothgang et al.~\cite{RRB23ext} also define a translation from DHOL to regular HOL, utilizing partial
equivalence relations to capture information that would otherwise be lost while erasing the
dependency in the types.

One of the changes to Church's original formulation is the omission of the choice operator. Choice is an 
important component of several higher-order interactive theorem provers such as Isabelle/HOL~\cite{NPW02}, 
HOL Light~\cite{H09} or PVS~\cite{ROS98}, whose dependently-typed specification language make it
particularly noteworthy. Choice is therefore interesting to automated reasoning too, as
illustrated by the successes 
of Satallax~\cite{BB11,B12}. This suggests that DHOL-capable theorem provers like
Lash~\cite{BK22,NBK24} will eventually benefit greatly from the availability of choice.

\textit{Contributions:} We
provide the necessary groundwork for such future work by extending
DHOL with an indefinite choice operator \(\varepsilon\) and the corresponding additions to the translation. We match the 
completeness results for the translation as given by Rothgang et al.~\cite{RRB23ext} and give arguments for the soundness.
We implement the erasure(s) in an extended version of Lash, introduce 34  DHOL problems with choice, and evaluate type-checking and proving on these problems.
The remainder of the paper is structured as follows: Section~\ref{sec:pre} sets the stage for the extension, introducing 
HOL and DHOL as well as the erasure between them. Following that, in Section~\ref{sec:choice},
we present two possible variants for the extension to the language and translation and
subsequently prove them both to be complete. We introduce some DHOL problems and evaluate the
respective translations in Section~\ref{sec:case}.

\section{Preliminaries}\label{sec:pre}

First, we present a formulation of higher-order logic (HOL), closely following the formulation given in~\cite{RRB23ext} 
and~\cite{NBK24} to facilitate a simple extension to DHOL in the next subsection. Figure~\ref{fig:holSynRules} gives the 
grammar of the HOL syntax.

\begin{figure}[t]
\centering
\begin{tabular}{l c l r}
\(T, U\) & ::= & \(\circ\) | \(T,a:tp\) | \(T,x:A\) | \(T,t\) & \quad theories \\
\(\Gamma ,\Delta\) & ::= & \(\cdot\) | \(\Gamma, x:A\) | \(\Gamma, s\) & context \\
\(A,B\) & ::= & \(a\) | \(A\rightarrow B\) | \(o\) & types \\
\(t,u,v\) & ::= & \(x\) | \(\lambda x:A.t\) | \(tu\) | \(\bot\) | \(t \Rightarrow u\) | \(t =_A u\) | \(\forall x:A.t\) & terms \\
\end{tabular}
\caption{\label{fig:holSynRules}HOL Syntax}
\end{figure}

A theory \(T\) is a concatenation of the empty theory \(\circ\) and any number of (simple) base type declarations 
\(a:tp\), typed variable and/or constant declarations \(x:A\), and axioms. A context \(\Gamma\) similarly concatenates the 
empty context \(\cdot\) with typed variables, typed constant declarations, and axioms, but misses base type declarations. 

Aside from base types, there is no distinction in how theories and contexts are handled. The difference is mainly 
semantic: Theories \textit{declare} the constants, types, and axioms that hold while contexts \textit{assume} that there 
are some variables of a certain type, or that some proposition is true. As such, during proof search --- for example 
in~\cite{NBK24} --- the theory is static, while there might be changes to the context.

Types are either the built-in boolean base type \(o\), base types \(a\) as defined in the theory or function types 
\(A\rightarrow B\). We will use upper-case letters \(A,B,\dots\) to denote type variables. Lastly, terms are 
variables/constants \(x\), \(\lambda\)-abstractions \(\lambda x:A.t\), applications \(tu\) or the built-in boolean 
constant \(\bot\) denoting the false proposition extended with the boolean connectives \(t \Rightarrow u\), \(t =_A u\) 
and \(\forall x:A.t\). We will use the lowercase letters \(t, u, v, \dots\) as term variables. We write \(\neg t\) for 
\(t \Rightarrow \bot\), \(t \neq_A u\) for \(\neg (t =_A u)\), \(\top\) for \(\neg\bot\) as well as \(t \land u\) 
and \(t \lor u\) for \(\neg(t\Rightarrow \neg u)\) and \(\neg t \Rightarrow u\) respectively. Lastly, \(\exists x:A.t\) 
will stand for \(\neg \forall x:A.\neg t\).

Application associates to the left, i.e.~\((tu)v \mathrel{\widehat{=}} tuv\), and we drop the subscript in \(t =_A u\) if it is clear 
from the context or irrelevant. We write \(t[x_1/u_1,\dots ,x_n/u_n]\) for the simultaneous capture-avoiding substitution of 
variables \(x_i\) with terms \(t_i\). 

The system uses the following judgments: We write \(\vdashs T\; thy\) and \(\vdashs_T \Gamma\; ctx\) for well-formed 
theories and contexts respectively. The \(s\) in the superscript makes clear we are talking about simple types. \(\Gamma 
\vdashs_T A\; tp\) and \(\Gamma \vdashs_T t:A\) establish that \(A\) is a well-formed type and the well-formed term \(t\) 
has type \(a\). A special case of this judgment is \(\Gamma \vdashs_T u\). In this case, the well-formed term \(u\) is of 
type \(o\) and also provable from \(T\) and \(\Gamma\). Lastly, we write \(\Gamma \vdashs_T A\equiv B\) to say that types 
\(A\) and \(B\) are judgmentally equal. This is a trivial statement in HOL but will become much harder once we introduce 
dependent types in the next section. Note that the \(T\) in the subscript of the turnstile is only ever absent when the 
statement talks about the well-formedness of the theory. As such we will drop it in the remainder, as it is fairly clear 
from the context whether or not it is technically required.

\subsection{DHOL}
We now make two minute changes to the previously defined syntax to allow for dependent types: First, we replace the 
function type \(A\rightarrow B\) with its dependent version \(\Pi x:A.B(x)\). Now the second type \(B\) may depend on a 
term of type \(A\). In the case where \(x\) does not appear free in B, we will stick to the arrow notation. The second 
change introduces dependent base types. A user-defined base type can now take any number of arguments of other --- already 
declared --- types. This base type can then be instantiated into a concrete type with terms of the fitting types. See 
Figure~\ref{fig:dholSynRules} for the exact changes to the grammar.

\begin{figure}[t]
\centering
\color{black!55}
\begin{tabular}{l c l r}
\(T, U\) & ::= & \(\circ\) | \textcolor{black}{\(T,a:(\Pi x:A.)^*\; tp\)} | \(T,x:A\) | \(T,t\) & \quad theories \\
\(\Gamma ,\Delta\) & ::= & \(\cdot\) | \(\Gamma, x:A\) | \(\Gamma, s\) & context \\
\(A,B\) & ::= & \textcolor{black}{\(at_1\dots t_n\)} | \textcolor{black}{\(\Pi x:A.B\)} | \(o\) & types \\
\(t,u,v\) & ::= & \(x\) | \(\lambda x:A.t\) | \(tu\) | \(\bot\) | \(t \Rightarrow u\) | \(t =_A u\) | \(\forall x:A.t\) & theories \\
\end{tabular}
\color{black}
\caption{DHOL Syntax (changes are highlighted)}
\label{fig:dholSynRules}
\end{figure}
We call base types with an arity of 0 \textit{simple base types} in reference to non-dependently typed HOL, as the 
fragment of DHOL where all base types have arity 0 is just HOL. The judgments stay the same, but we now write \(\vdashd\) 
to differentiate between HOL and DHOL. It is easy to see that it is now required for theories and contexts to be ordered, 
as the well-typedness of terms down the line may depend on the assumptions and assertions given previously.

More importantly, type-equality judgments \(\Gamma \vdashd A \equiv B\) are now significantly harder. Consider a type 
\(A := at_1\dots t_n\) and a type \(A' := at_1'\dots t_n'\). To establish that \(\Gamma \vdashd A \equiv A'\) it is necessary to 
show that for all \(i\) s.t. \( 1 \leq i \leq n\) \(\Gamma \vdashd t_i =_{A_i[x_1/t_1,\dots ,x_{i-1}/t_{i-1}]} t'_i\) is 
provable, i.e. DHOL is an extensional type theory and as such type checking is undecidable.

\subsubsection{Erasure}
As DHOL is rather new, there is not much native ATP support for the logic yet. However, the increased expressivity
in itself is already valuable. Rothgang et al. introduced a sound and complete translation from DHOL to HOL, thereby
providing automation support by serving as an interface between the DHOL language and any higher-order capable theorem
prover. We give an overview of the translation and refer to the original paper~\cite{RRB23ext} for the details and proofs.

The basic idea of the translation \(\overline{\square}\) is to reduce dependent types to their simple components and to 
encode the lost information in a partial equivalence relation (PER). In accordance to the
established notation, the PER generated while erasing a type \(A\) is written as \(A^*\).
The distinguishing feature of a PER is that it does 
not require reflexivity. Elements of a DHOL-type \(x:A\) are just those elements for which a PER \(A^*\) is indeed 
reflexive, i.e. \(\Gamma \vdashd t:A\) holds if and only if \(\overline{\Gamma} \vdashs A^*~\overline{t}~\overline{t} 
{\mbox{ and }} \overline{\Gamma} \vdashs \overline{t}:\overline{A}\).

Formally, the translation from DHOL to HOL is defined inductively on the grammar: 
\begin{align*}
\overline{\circ} \:&=\: \circ & \overline{\cdot} \:&=\: \cdot \\
\overline{T, U} \:&=\: \overline{T}, \overline{U} & \overline{\Gamma, \Delta} \:&=\: \overline{\Gamma}, \overline{\Delta} \\
\overline{o} \:&=\: o &
\overline{a t_1\dots t_n} \:&=\: a \\
\overline{\dpi x\colon A.B} \:&=\: \overline{A} \to \overline{B} &
\overline{x} \:&=\: x \\
\overline{\lambda x \colon A.t} \:&=\: \lambda x\colon\overline{A}.\:\overline{t} &
\overline{t\: u} \:&=\: \overline{t}\: \overline{u} \\
\overline{\bot} \:&=\: \bot &
\overline{\lnot t} \:&=\: \lnot \overline{t} \\
\overline{t \limp u} \:&=\: \overline{t} \limp \overline{u} &
\overline{t =_{A} u} \:&=\: A^*\: \overline{t}\: \overline{u} \\
\overline{\forall x\colon A.t} \:&=\: \forall x\colon\overline{A}.\: A^*\: x\: x \limp \overline{t} &
\overline{x\colon A} \:&=\: x\colon \overline{A}, A^*\: x\: x
\end{align*}
\begin{align*}
\overline{a\colon \dpi x_1\colon A_1.\:\cdots\: \dpi x_n\colon A_n.\:\tp} \:&=\: a\colon\tp \\
& a^*\colon \overline{A_1} \to \cdots \to \overline{A_n} \to a \to a \to o\\
& \text{with } \forall  x_1\colon\overline{A_1}.\: \dots \forall x_n\colon\overline{A_n}.\: \forall u,v\colon a.\: a^*\:x_1\:\dots\:x_n\:u\:v \limp u =_a v \\
o^*\: t\: u \:&=\: t =_o u \\
(a t_1\dots t_n)^*\: u\: v \:&=\: a^*\: \overline{t_1}\: \dots \: \overline{t_n}\: u\: v \\
(\dpi x\colon A.B)^*\: t\: u\:&=\: \forall x,y\colon\overline{A}.\: A^*\: x\: y \limp B^*\: (t\: x)\: (u\: y)
\end{align*}

\section{Choice}\label{sec:choice}

In order to allow choice in the input problems as well as in the inference rules,
we extend the HOL and DHOL term languages with $(\varepsilon x:A.t)$.
In HOL, the semantics of choice are more straightforward, since all types need to
be non-empty. This is not the case in DHOL, which is why we propose two variants of
typing rules for choice terms: the stronger $\varepsilon_1$ rule that requires the existence
of an element satisfying the predicate and the weaker $\varepsilon_2$ rule that only relies
on the type being inhabited. The mentioned inhabitation properties are stated by the
second premise in the according inference rules in
Figures~\ref{fig:dchoiceone}-\ref{fig:schoice}. Note that \(\Gamma,\forall x:A.\neg
t\vdash\bot\) is equivalent to \(\Gamma\vdash\exists x:A.t\).
These are in addition to the DHOL rules~\cite{RRB23ext}.

Remarkably, the prelude library of PVS~\cite{ROS98} defines two variants of choice function,
called \texttt{choose} and \texttt{epsilon}, that closely match our \(\varepsilon_1\) and
\(\varepsilon_2\) respectively. While we define two variants in order to investigate whether
one of those options is preferable, the fact that both formulations find use is noteworthy.

\begin{figure}[b!]
  \begin{gather*}
    \frac{\Gamma,x:A \vdashdchoiceone t:o \quad \Gamma,\forall x:A.\lnot t \vdashdchoiceone \bot}{\Gamma \vdashdchoiceone (\varepsilon x:A.t):A} \varepsilon_1 \text{\texttt{type}}
    \and\hspace{1cm}
    \frac{\Gamma,x:A \vdashdchoiceone t:o \quad \Gamma,\forall x:A.\lnot t \vdashdchoiceone \bot}{\Gamma \vdashdchoiceone t[x/(\varepsilon x:A.t)]} \varepsilon_1
  \end{gather*}
\caption{Choice Rules for DHOL${_{\varepsilon1}}$} 
  \label{fig:dchoiceone}
\end{figure}

\begin{figure}[b!]
  \begin{gather*}
    \frac{\Gamma,x:A \vdashdchoicetwo t:o \quad \Gamma,\forall x:A.\bot \vdashdchoicetwo \bot}{\Gamma \vdashdchoicetwo (\varepsilon x:A.t):A} \varepsilon_2 \text{\texttt{type}}
    \and\hspace{1cm}
    \frac{\Gamma,x:A \vdashdchoicetwo t:o \quad \Gamma,\forall x:A.\lnot t\vdashdchoicetwo \bot}{\Gamma \vdashdchoicetwo t[x/(\varepsilon x:A.t)]} \varepsilon_2 
  \end{gather*}
  \caption{Choice Rules for DHOL${_{\varepsilon2}}$} 
  \label{fig:dchoicetwo}
\end{figure}

\begin{figure}[b!]
  \begin{gather*}
    \frac{\Gamma,x:A \vdashschoice t:o}{\Gamma \vdashschoice (\varepsilon x:A.t):A} \text{\texttt{choice type}}
    \and\hspace{1cm}
    \frac{\Gamma,x:A \vdashschoice t:o \quad \Gamma,\forall x:A.\lnot t \vdashschoice \bot}{\Gamma \vdashschoice t[x/(\varepsilon x:A.t)]} \text{\texttt{choice}}
  \end{gather*}
\caption{Choice Rules for HOL\(_\varepsilon\)}
  \label{fig:schoice}
\end{figure}

The HOL rules are extended with choice in a straightforward way since simple types
are assumed nonempty.
Note the difference in formulation: DHOL\(_{\varepsilon1}\) has a stricter type formation rule --- requiring the ATP 
system to exhibit the existence of an element \(x\) such that \(t\) is true --- while DHOL\(_{\varepsilon2}\) only 
requires that the dependent type \(A\) is non-empty. The corresponding erasures respect this definition in that the 
stronger erasure assumes the provability of \(\exists x:A.t\) while the second and weaker erasure allows for a case 
distinction. The actual proof rule in either case requires the existence of a valid element of course.

Now that we can form well-typed terms in DHOL including choice and have specified what is provable,
we can we extend the erasure \cite{RRB23ext} to achieve automated proof support.
A first attempt might be to erase $(\varepsilon x:A.t)$ to the
simply typed term $(\varepsilon x:\overline{A}.\overline{t})$.
This erasure would make the translation ``incomplete.'' In particular,
we could have $\Gamma\vdashd_T (\varepsilon x:A.t) : A$
but not have $\overline{\Gamma}\vdashs_{\overline{T}} A^*~(\varepsilon x:{\overline{A}}.{\overline{t}})~(\varepsilon x:{\overline{A}}.{\overline{t}})$.
As a simple example, consider the DHOL theory $T$ given by $a:\dpi x:o.\; tp, c: a~\bot$
and the DHOL term $(\varepsilon x:a~\bot.\top)$.
Clearly we have $\cdot\vdashd_T (\varepsilon x:a~\bot.\top) : a~\bot$
in both the strong and (hence) weak senses.
However, $\cdot\not\vdashs_{\overline{T}} a^*~\bot~(\varepsilon x:a.\top)~(\varepsilon x:a.\top)$.

Consequently, we must somehow include information about $A^*$ in the erasure of $\varepsilon x:A.t$
(just as is done when erasing quantifiers).
In the case of the $\varepsilon$-operator, we need a different erasure depending
on whether we use the strong typing rule or the weak typing rule.
We will continue to use $\overline{t}$ for the strong erasure and introduce
the variant $\hat{t}$ for the weak erasure.
Likewise, we will continue to use the notation $A^*$ for the strong erasure of a type to a PER
and introduce the notation $A^{\hat{*}}$ for the weak erasure of a type to a PER.
The two erasures only differ in their treatment of $\varepsilon$.
The $A^*$ and $A^{\hat{*}}$ differ only in the sense of which erasure is used on
the terms of a dependent type.
That is, $(a~t_1\cdots t_n)^*$ is $a^*~\overline{t_1}\cdots\overline{t_n}$
while $(a~t_1\cdots t_n)^{\hat{*}}$ is $a^*~\hat{t_1}\cdots\hat{t_n}$.
All that remains is to specify
the values of $\overline{\varepsilon x:A.t}$
and $\widehat{\varepsilon x:A.t}$.
For the strong erasure we simply take
$\overline{\varepsilon x:A.t} = \varepsilon x:\overline{A}.A^*~x~x\land \overline{t}$.
For the weak erasure we need to consider two cases: when there is a witness satisfying
the predicate and when there is not. If there is a witness, we will use the
simply typed $\varepsilon$
just as in the strong erasure. However, when there is no witness, we will still
need to choose something of type $\overline{A}$ that will be related to itself by $A^*$.
If we had an if-then-else constructor in HOL, we could write $\widehat{\varepsilon x:A.t}$
as
$${\mathsf{if}}
~(\exists x:\hat{A}.A^{\hat{*}}~x~x\land\hat{t})~{\mathsf{then}}~(\varepsilon x:\hat{A}.A^{\hat{*}}~x~x~\land~\hat{t})~{\mathsf{else}}~(\varepsilon x:\hat{A}.A^{\hat{*}}~x~x).$$
As is well-known, if-then-else can be defined using $\varepsilon$ as 
$\lambda p x y. \varepsilon z. p \land z = x \lor \lnot p \land z = y$.
We remain in the HOL language we have already introduced,
and define $\widehat{\varepsilon x:A.t}$ to be
$$
\begin{array}{rl}
  \varepsilon z:\hat{A}.& (\exists x:\hat{A}.A^{\hat{*}}~x~x\land\hat{t})\land z = (\varepsilon x:\hat{A}.A^{\hat{*}}~x~x~\land~\hat{t})~\\
  & \lor\; \lnot (\exists x:\hat{A}.A^{\hat{*}}~x~x\land\hat{t})\land z =(\varepsilon x:\hat{A}.A^{\hat{*}}~x~x).
\end{array}
$$

\subsection{Completeness}
We will use \(\widetilde{\square}\) to denote either weak or strong
erasure, corresponding to the typing rule used. Using induction on the
structure of the natural deduction rules, we extend the completeness
proof given in \cite{RRB23ext} to get the following theorem:
\begin{thm}
For either variant --- DHOL\(_{\varepsilon1}\) or DHOL\(_{\varepsilon2}\) --- we retain that if \(\Gamma \vdashd t:A\) 
then \(\widetilde{\Gamma} \vdashs \widetilde{t}:\widetilde{A}\) and 
\(\widetilde{\Gamma} \vdashs A^*~\widetilde{t}~\widetilde{t}\). Also, if \(\Gamma \vdashd t\) then 
\(\widetilde{\Gamma} \vdashs \widetilde{t}\).
\end{thm}

\subsubsection{Proof of completeness for \(\varepsilon_1\) and strong erasure}

From the first assumption \(\Gamma, x:A \vdashd t:o\) we get the induction hypothesis \(\overline{\Gamma}, x:\overline{A}, 
A^*~x~x \vdashs \overline{t}:o\). With the \texttt{assume} rule, we can establish \(\overline{\Gamma}, x:\overline{A}, 
A^*~x~x \vdashs A^*~x~x\). Due to well-typedness, we know this is of type \(o\) and as HOL types cannot depend on 
propositions in the context, we may drop the \(A^*~x~x\) from the context in both hypotheses. Using \(\land\)-type on the 
hypotheses and a simple application of the definition of the strong erasure yields the first goal \(\overline{\Gamma} 
\vdashs \overline{(\varepsilon x:A.t)} : \overline{A}\). Note, that the first assumption of all
\texttt{type}-ing rules 
is the same, so this reasoning holds for either case. 

Next, we consider the second assumption \(\Gamma, \forall x:A.\neg t \vdashd \bot\) with the corresponding induction 
hypothesis \(\overline{\Gamma}, \forall x:\overline{A}.A^*~x~x \Rightarrow \neg \overline{t} \vdashs \bot\). By 
\(\Rightarrow_i\) and \(\neg\neg_i\) we arrive at \(\overline{\Gamma} \vdashs \neg \forall x:\overline{A}.
\neg\neg(A^*~x~x\Rightarrow\neg\overline{t})\) which is equivalent to \(\overline{\Gamma} \vdashs (\forall 
x:\overline{A}.\neg(A^*~x~x \land \overline{t})) \Rightarrow \bot\). We now put it back into the context and use the 
\texttt{choice} rule to conclude \(\overline{\Gamma} \vdashs (A^*~x~x \land \overline{t})[x/(\varepsilon x:
\overline{A}.A^*~x~x\land\overline{t})]\). The definition of the strong erasure applied to the term in the substitution 
and subsequent substituting of the \(x\) now yields our second goal and gives us the third one for free: 
\(\overline{\Gamma} \vdashs A^*~\overline{(\varepsilon x:A.t)}~\overline{(\varepsilon x:A.t)} \land \overline{t}
[x/\overline{(\varepsilon x:A.t)}]\). The derivation of \(\overline{\Gamma} \vdashs \overline{t}[x/\overline{(\varepsilon 
x:A.t)}]\) is valid because the \texttt{choice} rule in the \(\varepsilon_1\) case uses the same premises as the 
\texttt{choice type} rule.

\subsubsection{Proof of completeness for \(\varepsilon_2\) and weak erasure}

We now show the same results for the weak typing variant.
Since $\widehat{(\varepsilon x:A.t)}$ is defined using choice to implement if-then-else,
it is easy to prove that
$\Gamma\vdashs \widehat{(\varepsilon x:A.t)} = (\varepsilon x:\hat{A}.A^{\hat{*}}~x~x~\land~\hat{t})$
if $\Gamma\vdashs \exists x:\hat{A}.A^{\hat{*}}~x~x\land \hat{t}$.
Likewise,
if $\Gamma\vdashs \lnot\exists x:\hat{A}.A^{\hat{*}}~x~x\land \hat{t}$,
then
$\Gamma\vdashs \widehat{(\varepsilon x:A.t)} = (\varepsilon x:\hat{A}.A^{\hat{*}}~x~x)$.
We will use both facts implicitly below to rewrite ${\widehat{(\varepsilon x:A.t)}}$
according to one of these equations.

For both the \texttt{\(\varepsilon_2\)-type} rule and the \texttt{\(\varepsilon_2\)} rule
the inductive hypothesis of the first premise allows us
to infer
\(\widehat{\Gamma} \vdashs \widehat{(\varepsilon x:A.t)} : \hat{A}\)
as in the corresponding proof for the strong erasure.

We first show completeness for the \texttt{\(\varepsilon_2\)} rule.
The inductive hypothesis for the second premise
gives $\hat{\Gamma},\forall x:\hat{A}.A^{\hat{*}}~x~x\Rightarrow \lnot \hat{t}\vdashs \bot$.
From this we can derive
$\hat{\Gamma}\vdashs \exists x:\hat{A}.A^{\hat{*}}~x~x\land \hat{t}$
and
$\hat{\Gamma}, \forall x:\hat{A}.\lnot(A^{\hat{*}}~x~x\land \hat{t})\vdashs \bot$.
We need to prove 
$\hat{\Gamma}\vdashs \hat{t} [x/\widehat{(\varepsilon x:A.t)}]$.
It suffices to prove
$\hat{\Gamma}\vdashs (A^{\hat{*}}~x~x\land \hat{t}) [x/(\varepsilon x:\hat{A}.A^{\hat{*}}~x~x\land \hat{t})]$
which follows from the \texttt{simple choice} rule.

We now show completeness of the \texttt{\(\varepsilon_2\)-type} rule.
In this case the induction hypothesis of the second premise yields
$\hat{\Gamma},\forall x:\hat{A}.A^{\hat{*}}~x~x\Rightarrow \bot \vdashs\bot$.
We need to prove
$\hat{\Gamma}\vdashs A^{\hat{*}}~\widehat{(\varepsilon x:A.t)}~\widehat{(\varepsilon x:A.t)}$.
Unlike the previous argument, the inductive hypothesis is not sufficient to know 
the condition of the if-then-else is provable.
From here we proceed with a case distinction along the condition.
Since we are in a classical setting,
proving
$\hat{\Gamma},\exists x:\widehat{A}.A^{\widehat{*}}~x~x\land\hat{t} \vdashs A^{\hat{*}}~\widehat{(\varepsilon x:A.t)}~\widehat{(\varepsilon x:A.t)}$
and
$\hat{\Gamma},\lnot\exists x:\hat{A}.A^{\hat{*}}~x~x\land\hat{t} \vdashs A^{\hat{*}}~\widehat{(\varepsilon x:A.t)}~\widehat{(\varepsilon x:A.t)}$
suffice to prove
$\hat{\Gamma} \vdashs A^{\hat{*}}~\widehat{(\varepsilon x:A.t)}~\widehat{(\varepsilon x:A.t)}$.

We begin with the first case in which we assume \(\exists x:\widehat{A}.A^{\widehat{*}}~x~x\land\hat{t}\)
in the context.
Let $\Delta$ be $\hat{\Gamma},\exists x:\widehat{A}.A^{\widehat{*}}~x~x\land\hat{t}$.
Since $\Delta\vdashs \exists x:\widehat{A}.A^{\widehat{*}}~x~x\land\hat{t}$,
it suffices to prove
$$\Delta\vdashs A^{\hat{*}}~(\varepsilon x:\hat{A}.A^{\hat{*}}~x~x~\land~\hat{t})~(\varepsilon x:\hat{A}.A^{\hat{*}}~x~x~\land~\hat{t}).$$
By the \texttt{simple choice} rule, it suffices to prove
$\Delta,\forall x:\widehat{A}.\lnot (A^{\widehat{*}}~x~x\land\hat{t})\vdash\bot$.
This is obvious since \(\exists x:\widehat{A}.A^{\widehat{*}}~x~x\land\hat{t}\)
is literally the same as $\lnot\forall x:\widehat{A}.\lnot (A^{\widehat{*}}~x~x\land\hat{t})$.

Finally, we consider the second case in which the negation of the condition.
Let $\Delta$ be $\hat{\Gamma},\lnot\exists x:\widehat{A}.A^{\widehat{*}}~x~x\land\hat{t}$.
Since $\Delta\vdashs \lnot\exists x:\widehat{A}.A^{\widehat{*}}~x~x\land\hat{t}$,
it suffices to prove
$$\Delta\vdashs A^{\hat{*}}~(\varepsilon x:\hat{A}.A^{\hat{*}}~x~x)~(\varepsilon x:\hat{A}.A^{\hat{*}}~x~x).$$
By the \texttt{simple choice} rule, it suffices to prove
$\Delta,\forall x:\hat{A}.\lnot (A^{\hat{*}}~x~x)\vdashs \bot$.
This is precisely the inductive hypothesis of the second premise.

\subsection{Soundness}

Theorem 2 of~\cite{RRB23ext} gives a corresponding soundness result for the
translation from DHOL to HOL.
In particular, they prove $\Gamma\vdashd_T F$ holds whenever
$\Gamma\vdashd_T F:o$ and $\overline{\Gamma}\vdashs_{\overline{T}} \overline{F}$.
The proof of this result (found in the appendix of~\cite{RRB23ext})
is significantly more involved than the completeness
result. One complication is that the HOL proof of
$\overline{\Gamma}\vdashs_{\overline{T}} \overline{F}$
may make use of terms $t'$ that are not of the form $\overline{t}$
for a well-typed DHOL term $t$. (In~\cite{RRB23ext} these are called ``improper terms.'')
Semantics would provide an alternative strategy for proving soundness.
One could argue that if $\Gamma\not\vdashd_T F$, then there is a DHOL model
of $T$ satisfying $\Gamma$ but not $F$ (via a hypothetical completeness result for DHOL models).
Then one could argue that a DHOL model yields a Henkin model (for HOL)
of $\overline{T}$ satisfying $\overline{\Gamma}$
but not $\overline{F}$. Such a Henkin model would violate
soundness of HOL relative to Henkin models
since we are assuming $\overline{\Gamma}\vdash_{\overline{T}} \overline{F}$.
At the moment there is no accepted notion of DHOL model with soundness and completeness
results (even without choice), and no relationship between DHOL models and Henkin models
of HOL.
Consequently, for the present paper we leave soundness
(of both the weak and strong forms of DHOL with choice)
as a conjecture
and leave the sketch of a semantic argument as some indication why the
soundness results should hold.

\section{Experimental Problems and Results}\label{sec:case}

We propose several problems in DHOL that include choice and experiment with
the performance of the erasure on these problems. For the experiments, a modified version of Lash~\cite{BK22,NBK24} was run on the listed problems, once with each kind of erasure, for 90s.

The first set of problems share definitions for the type of natural numbers 
\(\mathsf{nat}\) with the corresponding constructors \(\mathsf{0}:\mathsf{nat}\) and \(\mathsf{s}:\mathsf{nat}\rightarrow 
\mathsf{nat}\), and the dependent-type of fixed-size finite sets \(\Pi n:\mathsf{nat}.\; tp\) with the two constructors 
\(\mathsf{fz}: \Pi n:\mathsf{nat}.\mathsf{fin}(\mathsf{s}\; n)\) and \(\mathsf{fs}: \Pi n:\mathsf{nat}.\mathsf{fin}\; n 
\rightarrow \mathsf{fin} (\mathsf{s}\; n)\). We use numbers \(1,2,\dots\) to mean \(\mathsf{s}\; 0, \mathsf{s}\; (\mathsf{s}
\; 0), \dots\).
Following is a short description of the different classes of problems:

\begin{compactitem}
\item \texttt{choice\_def1} has in addition a type for predicates \(p\) for elements of \(\mathsf{fin}\; 2\) and an axiom 
that establishes the existence of an element \(x:\mathsf{fin}\; 2\) which fulfills the predicate. The conjecture expresses 
the definition of the choice operator in this setting: \(p(\varepsilon x:\mathsf{fin}\; 2.p\; x)\).

\item \texttt{choice\_def2} is a generalization of \texttt{choice\_def} from \(\mathsf{fin}\; 2\) to \(\mathsf{fin}\; n\) where \(n \in \mathbb{N}\).

\item \texttt{choice\_def3} finally generalizes the previous example to an arbitrary dependent type.
\item \texttt{choice\_eq1/2} establish that the \(\varepsilon\)-operator respects identity for the types \(\mathsf{fin}\; 
1\), where there is only one element to choose from, and \(\mathsf{fin}\; 2\) respectively.
\item \texttt{choice\_nq} in turn, demonstrates that choice correctly chooses the other element in the \(\mathsf{fin}\; 
2\) type, or alternatively, that there are at least two elements in \(\mathsf{fin}\; 2\). 
\item \texttt{no\_fp\_finN\_reg} and \texttt{no\_fp\_finN\_min} is a family of problems for \(N
  \in \{0,\dots ,9\}\). Each 
has a conjecture that says that the function \(\lambda x:(\mathsf{fin}\; N).(\varepsilon y:(\mathsf{fin}\; N).x \neq y)\) has no fixed 
point. The variants with \texttt{\_reg} have \(\mathsf{fin}\; N\) behave like previously, while the \texttt{\_min} versions have 
fewer axioms and result in types that are assumed to have at least \(N\) elements as opposed to exactly \(N\) elements.
In case $N\geq 2$, it is clear that the term \(\lambda x:(\mathsf{fin}\; N).(\varepsilon y:(\mathsf{fin}\; N).x \neq y)\)
is well typed (of type $\mathsf{fin}\;N\to \mathsf{fin}\;N$) in both the strong and weak senses. Likewise it is provable that
the function given by the term has no fixed points.
When $N=1$, the term \(\lambda x:(\mathsf{fin}\; 1).(\varepsilon y:(\mathsf{fin}\; 1).x \neq y)\)
is well-typed in the weak sense (since $\mathsf{fin}\;1$ is provably nonempty)
but is not well-typed in the strong sense (since $\mathsf{fin}\;1$ does not provably have at least two elements).
When $N=0$, 
the situation depends on whether we have assumed $\mathsf{fin}\;0$ has precisely $0$ elements (a {\emph{provably}} empty type)
or at least $0$ elements (a {\emph{possibly}} empty type).
If $\mathsf{fin}\;0$ is provably empty, then
the term \(\lambda x:(\mathsf{fin}\; 0).(\varepsilon y:(\mathsf{fin}\; 0).x \neq y)\)
is well-typed in both the strong and weak sense, simply because
$\Gamma,x:\mathsf{fin}\; 0 \vdashd \bot$ allows us to prove
the premises of the relevant rules.
Likewise, if $\mathsf{fin}\; 0$ is provably empty, it is provable that the function has no fixed point
since quantification over an empty type is vacuous.
If $\mathsf{fin}\; 0$ is only possibly empty, then \(\lambda x:(\mathsf{fin}\; 0).(\varepsilon y:(\mathsf{fin}\; 0).x \neq y)\)
is not well-typed in the strong sense but is well-typed in the weak sense.
The reason it is well-typed in the weak sense is that,
although we cannot prove $\mathsf{fin}\;0$ is nonempty, we do
have $\Gamma,x:\mathsf{fin}\;0,\forall x:\mathsf{fin}\;0.\bot\vdashd \bot$.
\end{compactitem}

Additionally, several examples of problems related to fixed-length lists have been considered. These extend the 
definitions of natural numbers and successor function by definitions for the fixed length list, which has the same type as 
\(\mathsf{fin}\), \(\mathsf{nil}:\; \mathsf{list}\; 0\) and the constructor \(\mathsf{cons}: \Pi n:\mathsf{nat}.
\mathsf{nat}\rightarrow \mathsf{list}\; n\rightarrow \mathsf{list}\; (\mathsf{s}\; n)\).

\begin{compactitem}
\item \texttt{list\_empty} additionally introduces the predicate \(\mathsf{empty}\) that takes lists of any length and two 
axioms that ensure that the predicate is only satisfied when the list is indeed of length 0. The conjecture then 
establishes that choosing an empty list with \(\varepsilon\) satisfies the predicate.

\item \texttt{list\_nonempty} includes the same additional definitions but asserts that choosing a list of length 1 
does not satisfy the empty predicate.

\item \texttt{list\_head} introduces \(\mathsf{hd}: \Pi n:\mathsf{nat}.\mathsf{list}\; (\mathsf{s}\; n) \rightarrow 
\mathsf{nat}\) --- the unfailing head function. The conjecture asserts that the first element of a list chosen such that 
the chosen list's first element is 0, is indeed 0.
\end{compactitem}

\subsection{Results}\label{ss:res}

\begin{table}[tb!]
  \centering
  \begin{tabular}{lcccc}
    & \texttt{\(\varepsilon_1\) type-check} & \texttt{\(\varepsilon_1\) prove} & \texttt{\(\varepsilon_2\) type-check} & \texttt{\(\varepsilon_2\) prove} \\
    \texttt{choice\_def1..def3} & \cmark & \cmark & \cmark & \xmark \\
    \texttt{no\_fp\_fin\{0,2..9\}\_reg} & \cmark & \cmark & \cmark & \cmark \\
    \texttt{no\_fp\_fin\{0,2..8\}\_min} & \cmark & \cmark & \cmark & \cmark \\
    \texttt{no\_fp\_fin9\_min} & \cmark & \cmark & \cmark & \xmark \\
    \texttt{choice\_eq1} & \cmark & \cmark & \cmark & \xmark \\
    \texttt{choice\_eq2} & \cmark & \xmark & \cmark & \xmark \\
    \texttt{choice\_nq} & \xmark & \xmark & \xmark & \xmark \\
    \texttt{list\_empty} & \cmark & \cmark & \cmark & \xmark \\
    \texttt{list\_nonempty} & \xmark & \xmark & \xmark & \xmark \\
    \texttt{list\_head} & \cmark & \xmark & \cmark & \xmark \\
  \end{tabular} 
  \caption{Summary of the experimental results for the two typing/erasure variants.}
  \label{tab:results}
\end{table}

The results are presented in Table~\ref{tab:results}. 
Lash can easily type-check both
versions of all \texttt{choice\_def} problems. Proving the conjectures, however, only works well for the strictly typed
version --- requiring less than 10 seconds in all cases --- but times out when
using the weakly typed version. Similarly, \texttt{choice\_eq1} is 
type-checkable and provable with strong typing only. \texttt{choice\_eq2} still type-checks 
for both versions but is now provable by neither.
Lash is not able to type-check or prove \texttt{choice\_nq} for either of 
the choice typing variants.
The dependent list experiments are the hardest: Lash can only prove the simplest
of the three conjectures.

Lash does not type check the ill-typed problems among \texttt{no\_fp\_fin*} (included as a sanity test).
Lash also does not report them as provable.
With a 90s timeout Lash can type check and prove
each of the problems for $n \ge 2$ with one exception:
it does not prove \texttt{no\_fp\_fin9\_min}. The increased timeout becomes necessary as the
problems rise in difficulty.

We generated erased versions of the \texttt{no\_fp\_fin9\_min} problem,
and ran them on several other higher-order ATP systems. Only Leo-III~\cite{SB21} and
Zipperposition~\cite{BBTV23} were supporting native choice in the TPTP
THF language
and succeeded. Leo-III was able to prove the conjecture in 38s using the
strong erasure but timed out using the weak version. Zipperposition
took 0.02s for the strong erasure, and 0.96s for the weak one, suggesting
that the challenges posed by the weak typing are not Lash-specific.

Overall, surprisingly, type-checking performs comparably
well for weak and strong choice. However, using the strong typing rules results in
more than 50\% more problems proved, pointing towards favoring strong choice in
succeeding endeavors.

\section{Conclusion}

We have extended DHOL by dependent choice. Since types in DHOL can be empty,
we proposed two ways to specify the rules: a strong typing rule that ensures
that there exist witnesses and a weak choice typing rule that only checks that
the underlying type is non-empty. These are accompanied by a proof rule and
two erasure functions to enable automated reasoning. We also created a collection
of 34 dependent HOL problems that use/require dependent choice
(\url{http://cl-informatik.uibk.ac.at/cek/choice.tgz}). Our proofs and experiments show that
stronger choice works well with DHOL, making it the primary candidate for further
investigations. Future work includes implementation of native choice rules. 

\paragraph{Acknowledgements}
Supported by the Ministry of Education, Youth and Sports within the dedicated program ERC CZ under the project POSTMAN no. LL1902.
  This work has also received funding from the European Union’s Horizon
  Europe research and innovation programme under grant agreement
  no.~101070254 CORESENSE as well as the ERC PoC grant no.~101156734
  \emph{FormalWeb3}. Views and opinions expressed are however those of
  the authors only and do not necessarily reflect those of the
  European Union or the Horizon Europe programme. Neither the European
  Union nor the granting authority can be held responsible for them.

\bibliographystyle{plain}
\bibliography{references}

\appendix

\end{document}